\documentstyle[11pt,paspconf,epsf,twoside,makeidx]{article}
\makeindex
\begin{document}

\title{The PG X-ray QSO sample: Links between the UV-X-ray Continuum and Emission Lines}
\author{Beverley J. Wills\altaffilmark{1}, M. S. Brotherton\altaffilmark{2},
A. Laor\altaffilmark{3}, D. Wills\altaffilmark{1}, B. J. Wilkes\altaffilmark{4},
G. J. Ferland\altaffilmark{5}, \& Zhaohui Shang\altaffilmark{1}}

% Notice that some of these authors have alternate affiliations, which
% are identified by the \altaffilmark after each name.  The actual alternate
% affiliation information is typeset in footnotes at the bottom of the
% first page, and the text itself is specified in \altaffiltext commands.
% There is a separate \altaffiltext for each alternate affiliation
% indicated above.

\altaffiltext{1}{McDonald Observatory \& Astronomy Department, University of
Texas at Austin, TX 78712, USA }
\altaffiltext{2}{Institute of Geophysics \& Planetary Physics, Lawrence
Livermore National Laboratory, Livermore, CA 94550, USA }
\altaffiltext{3}{Department of Physics, Technion, Israel Institute of
Technology, Haifa 32000, Israel}
\altaffiltext{4}{Center for Astrophysics, 60 Garden Street, Cambridge MA 02138,
USA}
\altaffiltext{5}{University of Kentucky, Department of Physics and Astronomy,
Lexington, KY 40506, USA}

\begin{abstract}

The UV to soft X-rays of luminous AGNs dominate their bolometric luminosity,
driven by an accretion-powered dynamo at the center.  These photons ionize the
surrounding gas, thereby providing clues to fueling and exhaust.
Two sets of important relationships -- neither of them understood -- link the continuum
and gas properties.
%and properties of the gas on parsec to kiloparsec scales.
%and structure, dynamics and physical conditions of the broad emission line region (BLR)
%on sub-parsec and parsec scales, and the narrow line region (NLR) on kiloparsec scales.

\noindent (i) Boroson \& Green's `eigenvector 1' relationships:
Steeper soft X-ray spectra are clearly related to narrower H$\beta$ emission and 
stronger optical Fe\,II emission from the BLR, and weaker [O\,III]\,$\lambda$5007 from
the NLR.
We show that these relationships extend to UV spectra: narrower C\,III]$\lambda$1909,
stronger low ionization lines, larger
Si\,III]$\lambda$1892/C\,III]$\lambda$1909 (a density indicator), weaker 
C\,IV$\lambda$1549 but stronger higher-ionization N\,V$\lambda$1240.
We speculate that high accretion rates are linked to high columns of dense 
(10$^{10}$ -- $10^{11}$ cm$^{-3}$), nitrogen-enhanced, low-ionization gas from
nuclear starbursts.  Linewidth, inverse Fe\,II--[O\,III] and inverse Fe\,II--C\,IV
relationships hint at the geometrical arrangement of this gas.

\noindent (ii) The Baldwin effect (inverse equivalent width -- luminosity relationships):
Our correlation analyses suggest that these are independent of the above eigenvector\,1
relationships.  The eigenvector 1 relationships can therefore be used in future work, to
reduce scatter in the Baldwin relationships, 
%leading to further understanding of
%the connection between the luminous UV -- soft-X-ray continuum and nuclear material,
perhaps fulfilling the dream of using the Baldwin effect for cosmological studies.

\end{abstract}


\section{Introduction}

The UV to soft X-rays of luminous AGNs dominate their bolometric luminosity,
driven by an accretion-powered dynamo at the center.
Reprocessing of energy within a few gravitational radii determines the spectral
energy distribution at optical photon energies and above.  These photons illuminate and
ionize the surrounding gas on sub-parsec to kiloparsec scales -- the fuel and exhaust of
the central engine -- enabling us to investigate its structure, dynamics and physical
conditions.

Echo-mapping and photoionization modeling shows that luminous QSOs' strong broad
emission lines arise within 1 pc of the ionizing continuum, from gas with a range of
distances, velocities (1000 km s$^{-1}$ -- 10,000 km s$^{-1}$), densities (at least
$10^{9}$ cm$^{-3}$ -- $10^{12}$ cm$^{-3}$), and optical depths.  This is the broad
line region (BLR).  Its average
line strengths can be reproduced quite well (the ``locally optimally emitting
cloud'' model of Baldwin et al. 1995; see Korista's discussion in this volume).
Beyond the BLR, at distances up to many kiloparsecs, is lower density and lower
velocity ($\sim 500$ km s$^{-1}$) gas of the narrow line region (NLR).

Because QSO spectra are similar, the emphasis until recently, has been on explaining the
average QSO spectrum.  However, investigating {\it relationships} among QSO spectral
parameters (absorption and emission) as a
function of the properties of the ionizing continuum is a powerful way to further
explore how the central dynamo works (Francis et al. 1992).

There are two important sets of relationships linking central engine properties with the
surrounding gas  -- the Baldwin effect in the UV (Baldwin et al. 1978), and the 
optical--X-ray ``eigenvector\,1'' relationships first clearly presented by Boroson \&
Green (1992).

The Baldwin effect is an inverse relationship between QSO luminosity and the
equivalent width (EW) of C\,IV\,$\lambda$1549 (Baldwin et al. 1978), and other
broad emission lines (Kinney et al. 1990).  Because QSOs are exceedingly luminous
and therefore visible at high redshifts, EWs could serve to determine
luminosity distances.  QSOs could then be a tool for cosmology.  This hope has not
been fulfilled because there is too much scatter in the Baldwin relationships.

In optical QSO samples, much of the spectrum-to-spectrum variation is explained by the 
strong eigenvector 1 relationships (hereafter called principal component 1 or PC1): 
As H$\beta$ from the BLR becomes narrower, the strength of 
BLR Fe\,II (optical) emission increases, [O\,III] NLR emission decreases, and 
the optical--X-ray and X-ray spectra steepen (increasing
$\alpha_{ox}$ and $\alpha_x$, where F$_{\nu} \propto {\nu}^{-\alpha}$)
(Boroson \& Green 1992, Laor et al. 1994, 1997a, Grupe et al. 1998). 
These relationships link the emission-line properties to the central engine:\\
$\bullet$ The UV to soft X-ray region dominates the radiated power and ionizes the BLR.\\
$\bullet$ In principle, line widths (profiles), together with BLR distances based on
echo-mapping
time-delays or dust-sublimation radii, may be used to infer the virial central masses,
hence L$_{\rm Edd}$ (Peterson et al. 1998, Laor 1998).  QSO luminosities often appear to
be significant compared with L$_{\rm Edd}$, suggesting that accretion onto a central
Black Hole could be an unstable process, giving rise to winds.\\
$\bullet$ Line widths are also related to $\alpha_x$, linking kinematics directly to a
central engine property.  By analogy with Galactic Black Hole Candidates in their
``high'' states, steep X-ray spectra may be related to high accretion rates
(Pounds, Done, \& Osborne 1995).\\
$\bullet$ The structure of the BLR must be linked to $\alpha_x$ and hence probably to
accretion and outflowing winds.  One possible explanation for the inverse Fe\,II-[OIII]
relation is that the smaller the effective covering of the central engine by optically
thick BLR gas, the more photons are available to ionize the more distant NLR (Boroson
\& Green 1992, Brotherton, this volume).\\
$\bullet$ Clues to the geometry may also come from the fact that the BLR appears to see
more
continuum power than we infer from the observed continuum;
there are too few ionizing photons to explain the line
strengths -- especially low ionization lines like Fe\,II, but also He\,II$\lambda$1640
(Netzer 1985, Korista et al. 1997).

Here we present new UV relationships linking QSOs' spectral energy
distribution and the kinematics, structure and physical properties of the
emitting gas.  We show how these relationships could be used to reduce the scatter
in the Baldwin relationship and lead to understanding of the effect itself.
%use of BE for cosmology on firmer ground.

We have investigated the UV spectra of 22 QSOs\footnote{One object was
omitted for non-scientific reasons, so this omission does not bias the sample.}
from the complete, optically selected
sample of 23 QSOs with complete and high quality ROSAT (0.15 - 2 keV) spectroscopy
(Laor et al. 1994).  This is the
same sample for which Laor et al. (1994, 1997a) showed striking X-ray-optical
PC1 relationships.  A briefer report of the present work appeared in Wills et al. (1998a).

\section{ Observations \& Measurements}

\begin{figure}
\plotone{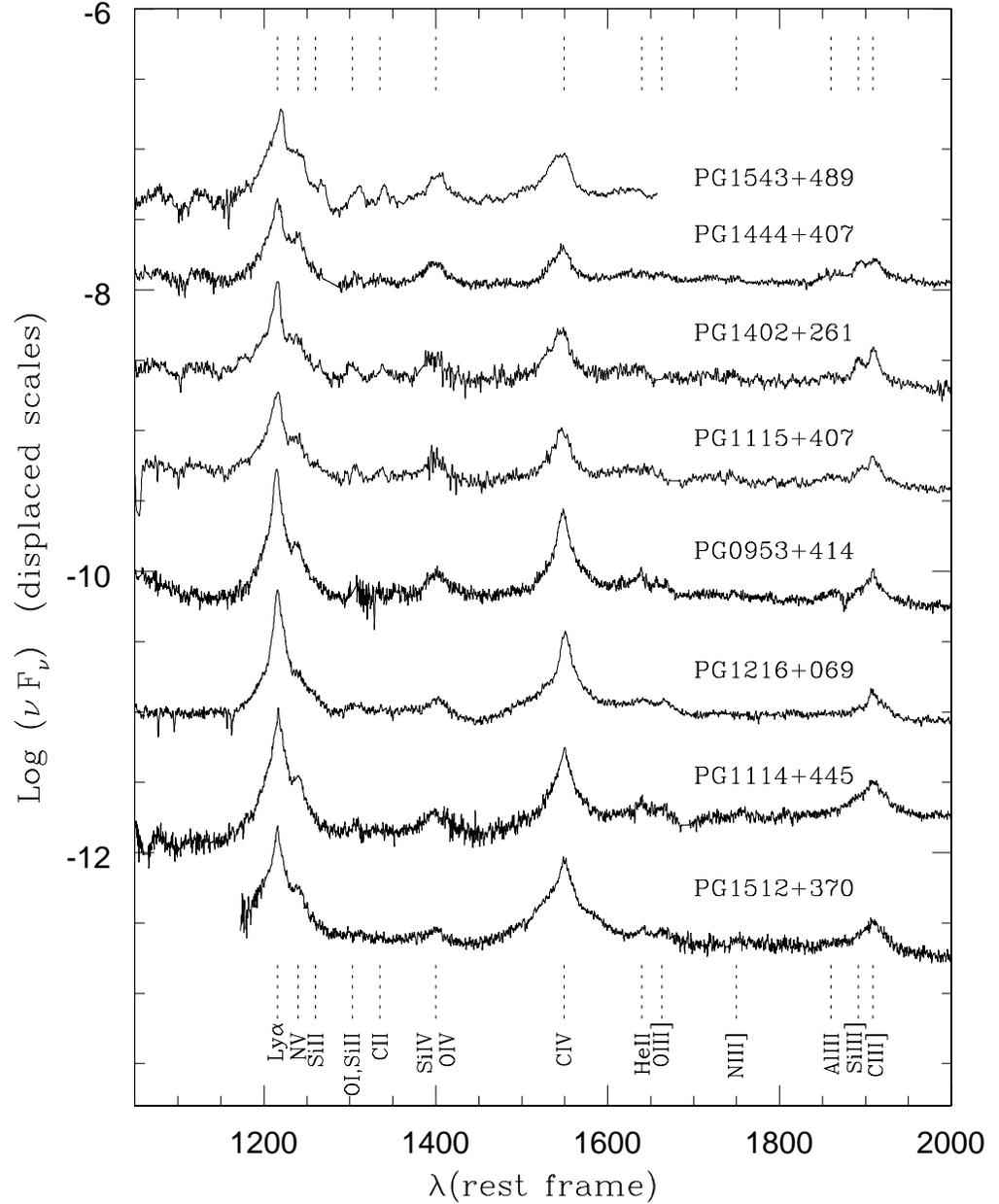}
\caption{
HST-FOS spectra of PG QSOs, plotted on a logarithmic flux density scale
to emphasize the weaker features.  They are plotted in order of PC1, with
the 3 lower QSOs having amongst the flattest X-ray spectra (smaller $\alpha_x$)
and weakest  Fe\,II(optical)),
and the 3 top spectra being those QSOs with softest X-ray spectra, strongest optical
Fe\,II(optical), weaker [OIII]$\lambda$5007, and narrowest (BLR) H$\beta$.  Increasing 
upwards, notice the increasing prominence of Al\,III$\lambda$1860, 
Si\,III]$\lambda$1892, O\,I$\lambda$1304, C\,II $\lambda$1335.
The $\lambda$1400 blend of O\,IV] and Si\,IV, as well as the apparent strength of 
NV\,$\lambda$1240 increase relative to CIV and Ly$\alpha$.
The strength of the optical and UV Fe\,II blends (not shown) also increases.
}
\end{figure}

We have obtained HST FOS spectrophotometry from wavelengths below Ly$\alpha$ to
beyond the atmospheric cut-off, and McDonald Observatory spectrophotometry
from the atmospheric cut-off to beyond H$\alpha$.  Instrumental resolutions range from
230 -- 350 km s$^{-1}$\,(FWHM).
Here we emphasize the new results from UV spectroscopy, with most optical data taken
from Boroson \& Green (1992).
In Figure 1 we show sample UV spectra, with QSOs arranged from the bottom in order of
increasing optical PC1, in the sense of steeper soft X-ray spectrum and
increasing strength of Fe\,II(optical).

We have measured strengths, ratios and widths (FWHM) for the following
emission lines of the BLR:
We have deblended Ly$\alpha$ and N\,V$\lambda$1240 from each other
and from Si\,II$\lambda$1260, O\,I$\lambda$1304 and C\,II$\lambda$1335.  We
measured C\,IV\,$\lambda$1549 with N\,IV\,$\lambda$1486,
He\,II\,$\lambda$1640 and [O\,III]\,$\lambda$1663 removed, and we have deblended
Al\,III$\lambda$1860, Si\,III] $\lambda$1892 and C\,III]\,$\lambda$1909.   In most
cases we used our McDonald spectra to define a `rest frame' wavelength scale referred to
[O\,III]\,$\lambda$5007 from the NLR.
The $\lambda$1909 feature is a blend of Si\,III$\lambda$1892, C\,III]$\lambda$1909 and 
Fe\,III lines.  For our sample Fe\,III is a minor contributor.
%  Generally, Fe\,III does not
%contribute much to the $\lambda$1909 blend.  
Evidence for this is that the
wavelength of the peak corresponds to within 0.5-1\AA\ rms of the expected
wavelength of C\,III]\,$\lambda$1909.
An exception is Mkn 478, where Fe\,III is a clearly present.
We have deblended the Ly$\alpha$--N\,V emission lines in several ways, some assuming the
N\,V
components to have the same width as Ly$\alpha$, and others assuming the same width 
as the components
of C\,IV$\lambda$1549, with qualitatively similar results.
The greatest uncertainties in line measurements arise from uncertainties in
continuum placement, and in removal of associated and Galactic interstellar
absorption.  Details will be presented by Wills et al. (1998b).  Some actual line
measurements are tabulated by Francis \& Wills (this volume).

\section{Correlation Results}
\index{eline!hbbeta}
\index{eline!oiii}
\index{eline!feii}
\index{eline!nv}
\index{eline!lya}
\index{eline!ciii}
\index{eline!civ}

\subsection{Direct Correlations}
\small

\begin{table}
\caption{Some Correlation Coefficients\tablenotemark{a}} \label{tbl-1}
\begin{center}
\begin{tabular}{rccccccc}
\hline
\\
%\vspace{-14pt}
%\footnotesize
UV Parameters & \multicolumn{6}{c}{Optical First Principal Component Parameters} & L$_{1216}$
\\
                       & $\alpha_x$
                       & log\,FW
                       & log\,EW
                       & Fe\,II/
                       & log\,EW
                       & opt-Xray\\
                        & 
                        & H$\beta$
                        & Fe\,II
                        & H$\beta$
                        & [O\,III]
                        & PC1\\
\tableline
\\
log\,FW\,C\,III]            & \bf $-$0.60  &   \bf +0.78  &  \bf $-$0.49  &  \bf $-$0.52
 &  +0.21  &  \bf $-$0.57 &  +0.18  \\
log\,EW\,Ly$\alpha$         &     $-$0.13   &  +0.27        &  $-$0.02        &  $-$0.16
 &  +0.34  &  $-$0.17 & \bf $-$0.77 \\
log\,EW\,C\,IV              & \bf $-$0.49  &   \bf+0.69  & \bf $-$0.53      &  \bf$-$0.62
 &   \bf+0.67  &  \bf $-$0.68 & $-$0.39 \\
C\,IV/Ly$\alpha$          & \bf $-$0.61  &   \bf+0.68  &  \bf $-$0.67  &  \bf$-$0.67
 & \bf +0.59  &  \bf $-$0.77 & $+$0.08 \\
log\,EW\,CIII]            &    $-$0.32  &     +0.42  &      $-$0.18  &  $-$0.36
 &  +0.41  &  $-$0.34 & \bf $-$0.68 \\
Si\,III]/C\,III]               & \bf $+$0.52 & \bf $-$0.56  &   \bf +0.52  &  \bf +0.89
 & \bf $-$0.62  &  \bf +0.62 & $-$0.10 \\
Si\,III/Ly$\alpha$          & $+$0.39       &  \bf $-$0.59  &  \bf+0.57   &  \bf +0.83
 &  \bf $-$0.70  &  \bf +0.72 & $-$0.22 \\
N\,V/C\,III]              & \bf $+$0.71 &  \bf $-$0.53  &  \bf +0.47  &  \bf +0.65
 & \bf $-$0.51  &  \bf +0.73 & +0.08 \\
N\,V/Ly$\alpha$           & \bf $+$0.53 & \bf $-$0.48     &   \bf +0.41  &  \bf +0.45
 & \bf $-$0.57  &  \bf +0.66 & $-$0.01 \\
$\lambda$1400/Ly$\alpha$  & \bf $+$0.69 &  \bf $-$0.60  &  \bf +0.60  &  \bf +0.58
 & \bf $-$0.57  &  \bf +0.71 & $-$0.31
\end{tabular}
\end{center}
\tablenotetext{a}{The usual Pearson correlation coefficients are given.\\
For 22 QSOs, a correlation coefficient of
0.4 corresponds to 1 chance in 15 of arising from uncorrelated variables
(two tailed),
0.5 corresponds to 1 chance in 50, % of arising from uncorrelated variables.\\
0.6 to 1 chance in 300, and % of arising from uncorrelated variables.\\
0.7 to 1 chance in 2000 of arising from uncorrelated variables.
FW $\equiv$ FWHM.\\
}
\end{table}

\normalsize

Some of the most important correlation results are summarized in Table 1 and Figure 2.
Column (1) of Table 1 lists emission-line parameters of the ultraviolet spectrum:
line ratios, logarithms of rest-frame equivalent widths (EW) and line widths
(full width at half maximum, FWHM or FW).  Columns (2) -- (6) give correlation 
coefficients
for the ultraviolet parameters vs. the well-known PC1 parameters, $\alpha_x$ derived
from a fit between 0.2\,keV and 2\,keV, and parameters of the optical spectrum as given
by Boroson \& Green (1992).  Column (7) gives correlation coefficients between the UV
parameters of col. (1) and a linear combination of X-ray--optical PC1 parameters
derived from
a principal components analysis of our sample (See \S3.2 and Francis \& Wills, this volume).
The table notes give two-tailed significance levels for the correlation coefficients.
Where the probability of a correlation arising by chance from
unrelated variables is $<$1 in 50, the coefficients are given in bold type.
\begin{figure}
\plotone{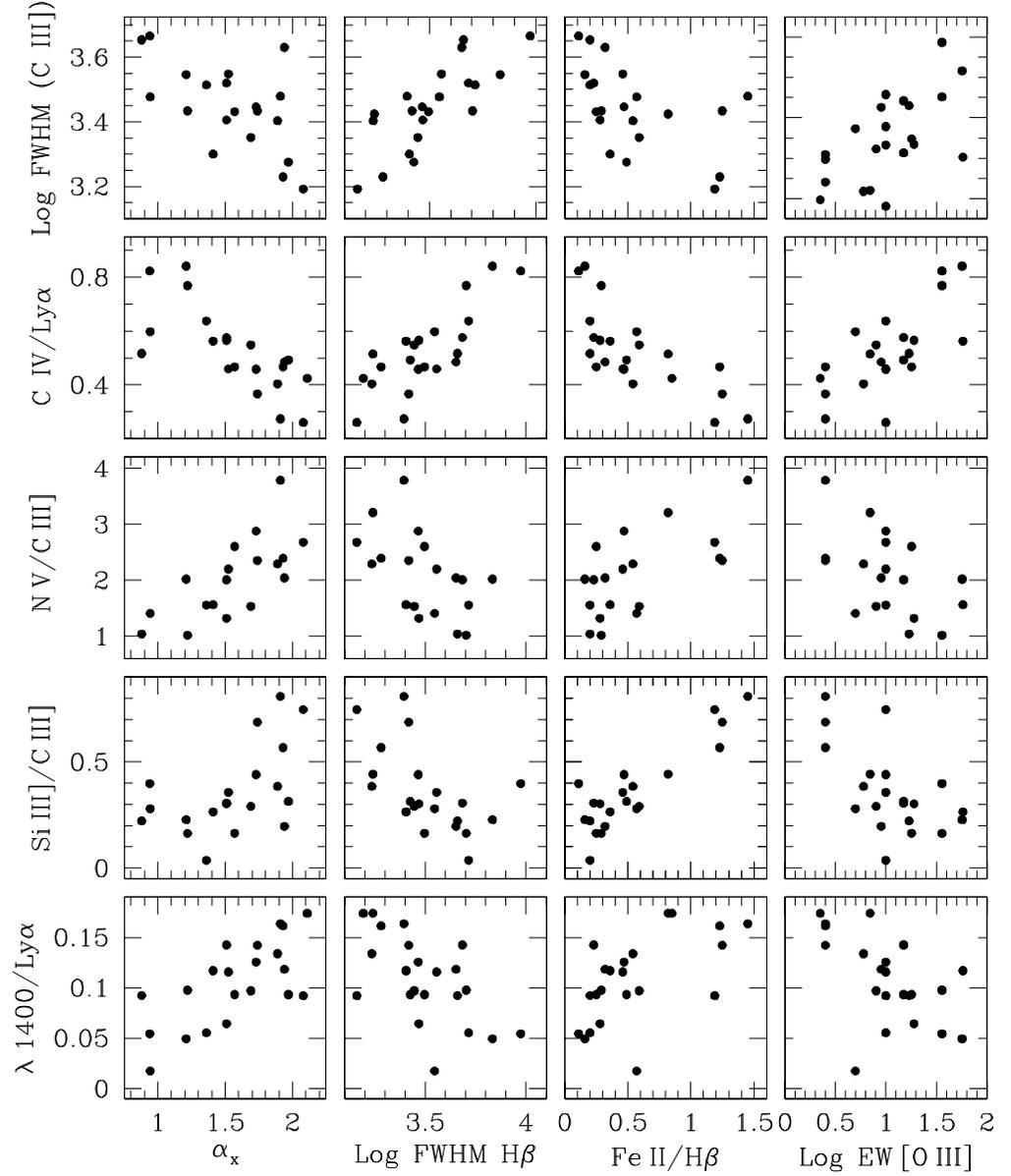}
\caption{UV observables vs. $\alpha_x$ and optical Eigenvector 1 observables.
Many line ratios are consistent with these 
trends -- lower ionization, higher densities, in addition to
(not shown) weaker [O\,III]$\lambda$5007 from the narrow line 
region (NLR) -- all corresponding to steeper soft X-ray
spectra.  The 2-tailed probability of these correlations
arising by chance from unrelated variables is between
1 in 50 to 1 in $>>$ 1000.}
\end{figure}
However, because such
a large fraction of our observationally-independent parameters are significantly
correlated, the assumption that the variables are unrelated breaks down, so
the significance of our correlations is actually higher than indicated in the table.
Figure\,2 plots some of the correlations of Table\,1,
the four columns representing X-ray--optical PC1 observables:
the steepness of the X-ray spectrum, the width of the broad H$\beta$ line,
the strength of Fe\,II(optical), and the strength of NLR emission
([O\,III]\,$\lambda$5007).
Optical PC1, in the sense of softer X-ray spectrum, clearly extends to
the UV: narrower C\,III] emission, stronger N\,V$\lambda$1240,
Si\,III]$\lambda$1892, and $\lambda$1400 feature.  Fe\,III UV\,34
($\lambda\lambda$ 1895, 1914, 1926) also contributes more when Fe\,II is strong.
Mkn\,478 in our sample, and I\,Zw\,1 (see Laor et al. 1997b), have especially soft
X-ray spectra, especially strong Fe\,II(optical), as well as strong Fe\,III (I\,Zw\,1
did not meet the Galactic obscuration criteria for inclusion in the sample, but is
otherwise eligible).
The strength of C\,IV$\lambda$1549 is inversely correlated with the strength of
Fe\,II (optical) (Wang et al. 1996) and correlated with [O\,III] 
(Brotherton, this volume.)
There are also very significant correlations among the UV parameters by themselves
(\S3.2).
Correlations as strong as we have found should show up in a visual inspection of the
spectra.  They do (see Figure 1, being careful to account for the effects of
line width on blended features).

\begin{figure}
\plotone{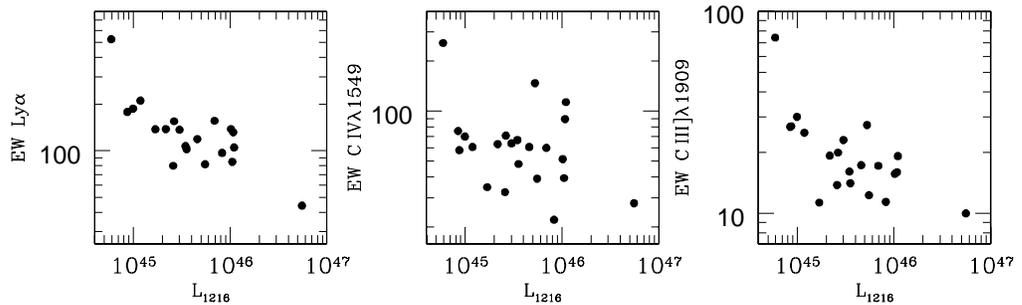}
\caption{The Baldwin relationships for Ly$\alpha$, C\,IV$\lambda$1549 and C\,III]$\lambda$1909
for $\sim$20 QSOs of the PG X-ray  sample.}
\end{figure}

Also shown in Table 1 and Fig. 3 are correlations between UV emission-line parameters
and L$_{1216}$, log of the continuum luminosity at 1216\AA.  The Baldwin relationships are
present.  The significance  depends largely on the
highest luminosity QSO 3C\,273 with the smallest EWs, and the lowest luminosity QSO
PG\,1202+281 with the largest EWs.  A larger sample is needed to show this result
clearly.

\subsection{Principal Components Analysis}
\small

\begin{table}
\caption{Principal Components Analysis} \label{tbl-2}
\begin{center}
\begin{tabular}{rrrr}
\tableline
\\
              &   PC1    &  PC2    &  PC3       \\
\tableline
 Eigenvalue   &  6.49   & 2.47   & 1.63   \\
 Proportion   &   0.499   &  0.190   &  0.126     \\
 Cumulative   &   0.499   &  0.689   &  0.815      \\
\\
 Variable     &   PC1    &  PC2    &  PC3       \\
\tableline

 L$_{1216}$                & $+$0.01  & $+$0.52  & $-$0.30    \\
 $\alpha_x$                & $+$0.32  & $-$0.16  & $+$0.03    \\
 FWHM H$\beta$             & $-$0.35  & $+$0.03  & $-$0.32    \\
 Fe\,II/H$\beta$           & $+$0.35  & $-$0.09  & $+$0.09    \\
 EW [O\,III]               & $-$0.30  & $+$0.02  & $+$0.25    \\
 FWHM C\,III]              & $-$0.20  & $-$0.06  & $-$0.61    \\
 EW Ly$\alpha$	           & $-$0.15  & $-$0.51  & $+$0.10    \\
 EW C\,IV                  & $-$0.33  & $-$0.24  & $+$0.05    \\
 C\,IV/Ly$\alpha$          & $-$0.34  & $+$0.18  & $+$0.03    \\
 EW C\,III]                & $-$0.25  & $-$0.47  & $-$0.08    \\
 Si\,III]/C\,III]          & $+$0.35  & $-$0.06  & $-$0.02    \\
 N\,V/Ly$\alpha$           & $+$0.23  & $-$0.14  & $-$0.54    \\
 $\lambda$1400/Ly$\alpha$  & $+$0.23  & $-$0.31  & $-$0.24
\end{tabular}
\end{center}
\tablenotetext{}{There were 18 QSOs used in this analysis.
L$_{1216}$ represents continuum luminosity at 1216\AA.
}
\end{table}

\normalsize

We perform a principal components analysis (PCA) to investigate which sets of 
observational variables correlate or anticorrelate together, and therefore whether
we can define a smaller number of new variables that are linear combinations of 
the measured variables.   This would simplify a description of our dataset.  Moreover,
it may suggest one or more physical processes giving rise to the spectrum-to-spectrum
variations.  An introduction to PCA is given in Francis and Wills of this volume, where our
present dataset is used as an example.

\index{corr!eigen!uv}
We choose some of the same variables as as in \S3.1, and perform the analysis on the
ranks of the measurements for each observational variable.  The results are shown
in Table 2\footnote{For the results of a PCA on the unranked data, see Table 3,
of Francis and Wills in this volume).}. 
Half of the sample variance (49.9\%) is described by a linear combination
of variables related to the optical--X-ray principal component PC1 found by Boroson \&
Green (1992) and Laor et al. (1994, 1997a).
The weights of these measured parameters are given in the PC1 column.

\index{corr!eigen!uv}
The second most important principal component (PC2)
is dominated by an inverse relationship between L$_{1216}$ and EW.
All EWs except EW H$\beta$, EW [O\,III], and
EW Fe\,II(optical) show likely inverse relationships between EW and luminosity, and 
the correlation appears particularly strong for O\,VI$\lambda$1034 (not shown; see
the results of Zheng et al. 1995, for a large heterogeneous sample.)  These are the
same trends as seen in the classical Baldwin relationships (Kinney et al. 1990).
The most interesting result of the PCA is that PC1 appears not to depend on luminosity
at all (correlation coefficient $\sim$0.1):  The Baldwin relationships appear
independent of the X-ray--UV--optical PC1 relationships.  C\,IV$\lambda$1549 is
present in PC1 and PC2, probably accounting  for its weaker Baldwin relationship
in our sample (Fig. 3).

PC3 appears to represent an independent relationship between luminosity and line widths,
but we defer consideration of this until a larger sample is available.
Each of the remaining principal components account for less than 5\% of the sample
variance and are not significant.

\section{Summary}
We show that optical principal component 1, linking steeper soft X-ray spectra with
increasing optical Fe\,II strength, decreasing [O\,III]$\lambda$5007
emission, and narrower BLR H$\beta$ emission, extends to the UV emission lines:
narrower C\,III]$\lambda$1909, weaker C\,IV$\lambda$1549 emission, stronger 
low ionization emission lines -- UV Fe\,II, %C\,III]$\lambda$1909,
Si\,III]$\lambda$1892, and probably O\,I$\lambda$1305, C\,II$\lambda$1335,
and Fe\,III, and also a larger ratio
 Si\,III]$\lambda$1892/C\,III]$\lambda$1909.  
\index{corr!eigen!uv}But 
N\,V and the O\,IV--Si\,IV $\lambda$1400 blend also increase in strength with 
steeper soft X-ray spectrum, increasing Fe\,II(optical), decreasing H$\beta$ width
and decreasing [O\,III] strength.  
Figure 4 presents an approximate summary of these correlations and anticorrelations,
based on the results for many line ratios.

\begin{figure}
\plotone{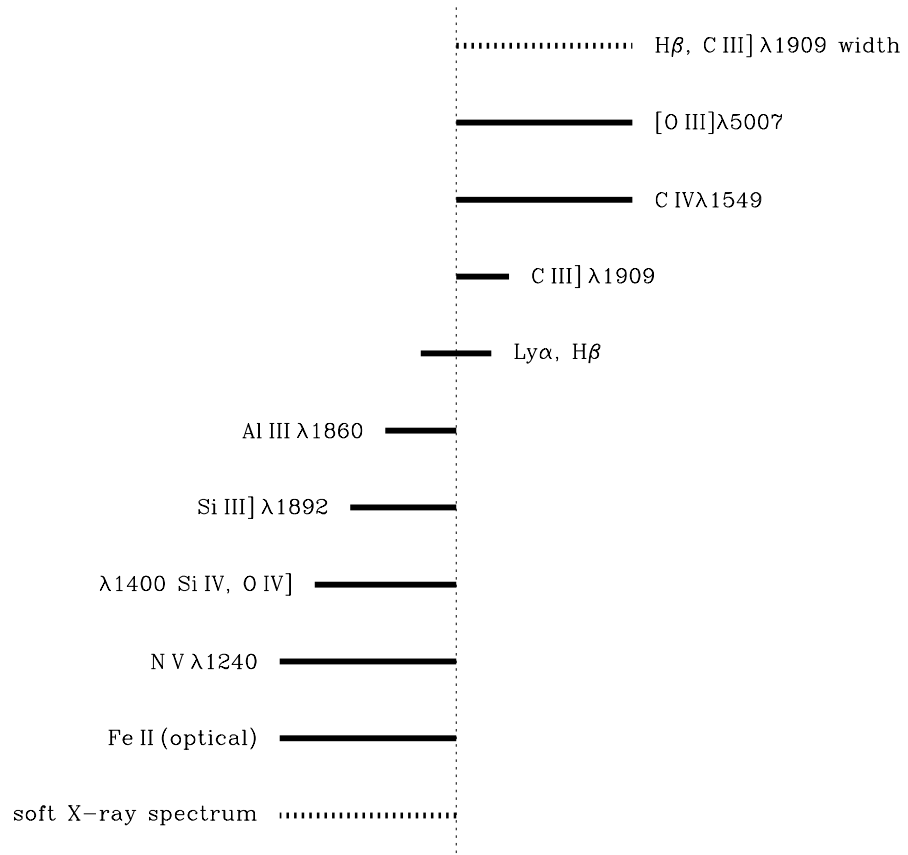}
\caption{An approximate description of the relationships among $\alpha_x$,
UV PC1 variables, and optical PC1 variables.
Variables on the same side of the center correlate positively with each
other, and on opposite sides, negatively.   The lengths of the  bars indicate
an approximate strength of the correlation.}
\end{figure}
While Fe\,II(optical) vs. $\lambda$1400/C\,IV and C\,IV/Ly$\alpha$, and 
$\alpha$(UV--X-ray)
vs. C\,IV/Ly$\alpha$ relationships have been noted before in large, heterogeneous
samples (Wang et al. 1996, 1998; see also Dultzin-Hacyan 1997), the spectacular
related correlations that we have found among many variables may be attributed to
the homogeneous and complete
nature of the sample.  For further illustration and discussion of the Fe\,II --
Si\,III]/C\,III] relationships, as well as Al\,III/C\,III] and Fe\,III/C\,III], see
Aoki \& Yoshida (this volume), and for an independent presentation of the
C\,IV -- [O\,III] relationship, see Brotherton (this volume).

\section{Discussion}

In the Introduction, we showed that the X-ray--{\it optical} relationships provide 
plausible strong links between central engine parameters and 
%the kinematics, geometry and physics of the 
ionized gas on sub-parsec
to Kpc scales.  Clearly the UV--soft-Xray continuum photoionizes the
surrounding gas and the emission lines can, in principle, provide information on the
shape of the ionizing continuum.  {\it UV spectrophotometry} provides further pieces for the
puzzle by relating the central engine to the kinematics, geometry and density of
fueling gas or outflows:\\
$\bullet$ Si\,III]$\lambda$1892 probably arises from almost the same region as
C\,III] and most optical and UV
Fe\,II, but C\,III]$\lambda$1909 is collisionally suppressed at high densities
$>$10$^{10}$ cm$^{-3}$, thus increasing the ratio Si\,III/C\,III] (e.g. Laor et al. 
1997b).  Our observed dependence of Si\,III]/C\,III] on Fe\,II strength is therefore
consistent with the high densities and column densities required to produce strong
Fe\,II(optical) (Ferland \& Persson 1989, Wills, Netzer \& Wills 1985).\\
$\bullet$ The C\,IV emission is strongly correlated with [O\,III]$\lambda$5007.  As
suggested to explain the inverse Fe\,II-[OIII] relation, does this mean that 
thick high density Fe\,II-emitting gas shields more distant C\,IV-emitting gas?\\
$\bullet$ The strength of the higher-ionization N\,V$\lambda$1240 also appears to be
associated with this higher density gas, while C\,IV strength decreases with increasing
PC1 (i.e., Fe\,II, etc.)
This apparent anomaly is reminiscent of the behavior of the EW(N\,V) Baldwin
relation in which the slope of the Baldwin relation for
several other species is tightly correlated with ionization potential, but for
N\,V the dependence on luminosity is much less than expected.  Korista et al.
(this volume, 1998) show that the EW (N\,V) -- luminosity relationship can be understood if
very N-rich gas (Hamann \& Ferland 1993) is irradiated by a UV--soft-Xray continuum whose
strength increases with luminosity.  While this is a complex issue, perhaps our N\,V
anomaly can be similarly explained by enhanced N abundances.

\index{corr!eigen!uv}
The `mystery physics' underlying the PC1 relationships could be an increase in
L/L$_{\rm Edd}$ associated with an increase in dense, Fe\,II emitting gas of
high metallicity.
It has been suggested that the extreme low ionization spectra with strong Fe\,II are
somehow related to starbursts (L\^{\i}pari et al. 1993).  Could starburst
activity cause high accretion rates, hence strong soft X-rays?

What do our new results mean for the Baldwin Effect?
Our PCA suggests that the strong first principal component (PC1) relationships
that we find
are independent of luminosity.  The second principal component, PC2, apparently
represents the Baldwin effect -- inverse relationships between EW and luminosity.
Because we find strong links among line strengths and UV--X-ray continuum
parameters, it is clear that these relationships must cause significant variation in
equivalent widths from one QSO to another.
Even if their cause is unknown, 
the relationships between line strengths and observables independent of L could be 
used to remove most of the scatter in Baldwin relationships.
Could the cause of PC1 remove not only the scatter from
the Baldwin relationships, but also remove the Baldwin effect itself?
Not if PC1 is independent of luminosity; investigation of larger samples is needed.
However, if the variance of
emission-line strengths could be attributed entirely to differences in the ionizing
spectral energy distribution, the crucial question would then become: How and why
does the ionizing spectral energy distribution
depend on luminosity?, and this gets to the heart of the Central Engine.

\acknowledgments

We gratefully acknowledge the help of 
C. D. Keyes \& A. Roman of STScI, and M. Dahlem (now of ESTEC),
also M. Cornell and R. Wilhelm who provide computer support in
the Astronomy Department of the University of Texas.
This research is supported
by NASA through LTSA grant number NAG5-3431 (B.J.W.) and grant number
GO-06781 from the Space Telescope Science Institute, which is operated by
the Association of Universities for Research in Astronomy, Inc., under
NASA contract NAS5-26555.

\end{document}